\documentclass{elsart1p}
\usepackage{amsfonts,amssymb,amsmath,graphics}
\usepackage{graphicx}
\newcommand*{\dis}{\displaystyle}
\newcommand*{\bs}{\boldsymbol}

\begin{document}

\begin{frontmatter}
\title{\bf Full and Partial Thermalization of Nucleons in Relativistic
       Nucleus-Nucleus Collisions }

\author{D.~Anchishkin[1], A.~Muskeyev[2], S.~Yezhov[2]}

\address{[1]{Bogolyubov Institute for Theoretical Physics,
             03680 Kiev, Ukraine}}
\address{[2]{Taras Shevchenko Kiev National University,
03022 Kiev, Ukraine}}
\begin{abstract}
We propose a mechanism of thermalization of nucleons in
relativistic nucleus-nucleus collisions. Our model belongs, to a
certain degree, to the transport ones; we consider the evolution
of the system, but we parametrize this development by the number
of collisions of every particle in the system rather than by the
time variable. We based on the assumption that the nucleon
momentum transfer after several nucleon-nucleon (-hadron)
collisions becomes a random quantity driven by a proper
distribution.
\end{abstract}

\end{frontmatter}

\noindent {\bf{The Model.}} \hspace{2mm} Our model is aimed at the
description of the nucleon spectra ($dN/dy$, $dN/m_{\bot} dm_{\bot}
dy$) for such collision energies, when the number of created
nucleon-antinucleon pairs is much less than the number of net
nucleons, i.e. it can be applied at AGS and low SPS energies.

We now make three key assumptions about the nucleon system.

1. We separate all nucleons after freeze-out into two groups in
accordance with their origination: a) the {\it first group}
consists of net nucleons that went through hadron reactions; b)
the {\it second group} includes nucleons which were created in the
collective processes, for instance, during the hadronization of a
QGP. Then, the nucleon momentum spectrum can be represented as a
sum of two different contributions:
\begin{equation}
\frac {d N}{d^3p} = \left( \frac {d N}{d^3p} \right)_{\rm hadron} +
\left( \frac {d N}{d^3p} \right)_{\rm QGP} \,. \label{model1}
\end{equation}
In turn, the total number of registered  nucleons equals $N_{\rm
total} = N_{\rm hadron} + N_{\rm QGP}$. If this separation can be
done, we can define the ``nucleon power'' of the created QGP as
$P_{\rm QGP}^{(N)}=N_{\rm QGP}/N_{\rm total}$. In the present work,
we mainly deal with the nucleons from the first group.

2. The collision number for every nucleon (hadron) is finite
because the lifetime of the fireball is limited.
To determine the maximal number of collisions, $M_{\rm max}$, in
a particular experiment, we use
the results of UrQMD calculations \cite{urqmd1,urqmd2}.

3. Because the colliding nuclei are spatially restricted
many-nucleons systems, the different nucleons experience different
collision numbers: it is intuitively clear that the collision
histories of the inner and surface nucleons will be different.
That is why, we partition all amount of nucleons of the first
group (nucleons which take part in hadron reactions only) into
different ensembles in accordance with a number of collisions
before freeze-out. Then the nucleons from every ensemble give
their own contribution to the total nucleon spectrum. If we denote
the number of particles in a particular ensemble where the
particles experienced $M$ effective collisions by $C(M)$, then, in
correspondence to (\ref{model1}), we can write the total nucleon
spectrum as
\begin{equation}
\frac{d N}{d^3 p} = \sum_{M=1}^{M_{\rm max}} \, C(M) \,
D_M(\bs p) + C_{\rm therm} \, D_{\rm therm}(\bs p) \, ,
\label{model2}
\end{equation}
where $D_M(\bs p)$ is the spectrum (normalized to unity) of the
particles in the $M$-th ensemble. The last term on the r.h.s. of
(\ref{model2}) corresponds to the possible contribution from the
totally thermalized source which we associate with a QGP. Here,
$C_{\rm therm}$ is the number of nucleons which are created during
the hadronization of the QGP, and $D_{\rm therm}(\bs p)$ is the
thermal distribution.

Consider successive variations of the momentum of the $n$-th
nucleon from nucleus $A$ which moves with momentum ${\bs k}_0$
along the collision axis from left to right toward nucleus $B$.
Every $m$-th collision induces the momentum transfer
${\boldsymbol{q}} _n^{(m)}$ for the $n$-th nucleon. So that, after
$M$ collisions, the nucleon acquires the momentum
${\boldsymbol{k}}_n = {\bs k}_0+{\bs Q}_n$, where $\bs Q_n =
\sum_{m=1}^M \bs q_n^{(m)}$ is the total momentum transfer finally
obtained by the $n$-th nucleon.

We assume for the moment that the elastic scattering gives the main
contribution to the two-nucleon collision amplitude. The initial
momentum of every nucleon in nucleus $A$ is
$\boldsymbol{k}_a=\boldsymbol{k}_0=(0,0,k_{0z})$, while the initial
momentum of every nucleon in nucleus $B$ is
$\boldsymbol{k}_b=-\boldsymbol{k}_0=(0,0,-k_{0z})$, in the c.m.s. of
colliding nuclei. The energy and momentum are conserved in every
separate collision of two particles, $ \omega({\bs k}_a) +
\omega({\bs k}_b) = \omega({\bs p}_a) + \omega({\bs p}_b), ~~ \bs
k_a + \bs k_b  = \bs p_a + \bs p_b$, where $\boldsymbol{p}_a$ and
$\boldsymbol{p}_{b}$ are the momenta of the particles after
collision.
We assume that the particles are on the mass shell, so
that $\omega (\boldsymbol{k})=\sqrt{m^2+\boldsymbol{k}^2}$ (the
system of units $\hbar =c=1$ is adopted).
Determining the six unknown quantities, $\bs p_a$ and $\bs p_b$, from
four equations is straightforward but two quantities
(e.g., $(\boldsymbol{p}_a)_x$ and $(\boldsymbol{p}_b)_x$) remain
uncertain and can be considered as such which accept random values
driven by the scattering probability.
After the third collision, every component of the particle momentum
becomes completely random.
If the initial momentum is fixed, this means the full randomization of
the momentum transfer after three successive collisions.
So, if we follow
the elastic scattering of a nucleon from the first collision
to the last one, we would see the full randomization of the
momentum transfer after every three successive scattering acts.
As for the inelastic collisions, the nucleon momentum transfer
undergoes the even faster randomization \cite{anch-2008-1}.

First, we would like to determine the density distribution
function $f_{2N}$ in the momentum space, which describes $2N$
nucleons after $M$ collisions per particle. The whole consideration is
carried out in the c.m.s. of two identical colliding nuclei.
Let us write down a density distribution
function in the form $f_{2N}=C\widetilde{f}_{2N}$, where $C$ is
the normalization constant. The unnormalized distribution function
$\widetilde{f}_{2N}$ can be defined in a two-fold way: first, we
follow all collisions of a particular nucleon by the integration with
respect to all nucleon random momentum transfer, and, second, we
fix the total energy of the $2N$-nucleon system after freeze-out in a
microcanonical-like way:
$E_{\mathrm{tot}}=\sum_{n=1}^{2N}\omega (\boldsymbol{k}_n)$.
The integration measure of the momentum transfer for
the $n$-th particle in a series of $M$ collisions is determined as
\begin{equation}
dQ_n \equiv \prod_{m=1}^M \, J_m\left({\bs q}_n^{(m)}\right)\,
\frac{d^3q_n^{(m)}}{(2\pi)^3} \quad {\rm with} \quad \int
\frac{d^3q}{(2\pi)^3} \, J_m({\bs q})=1 \, , \label{0}
\end{equation}
where the distribution of the momentum transfer is characterized
by the presence of the form-factor $J_m({\bs q})$. For the sake of
simplicity, we assume the independence of $J_m(\bs q)$ on the
collision number $m$, i.e. $J_m({\bs q})\to J(\bs q)$.
Hence, we adopt an approximation where just one form-factor $J(\bs q)$
characterizes a distribution of the momentum-transfer in a series of
collisions which are experienced by a nucleon during its traveling
through the fireball.

Then the unnormalized
$2N$-particle distribution function reads
\begin{equation} \begin{array}{l}
\widetilde{f}_{2N}(E_{\rm tot};{\bs k}_1,\ldots , {\bs k}_{2N} )
=
\delta \! \left( E_{\rm tot} - \sum_{n=1}^{2N} \omega(\bs k_n)
\right) \int \frac{dQ_1}{V} \ldots \frac{dQ_{2N}}{V}
\\  \times
\prod_{n=1}^N \left[ (2\pi)^3 \delta^3 \! \left( {\bs k}_n- {\bs
k}_0 - \sum_{m=1}^M {\bs q}_n^{(m)}\right) \right]
\prod_{n=N+1}^{2N} \left[  (2\pi)^3 \delta^3 \left( {\bs k}_n +
{\bs k}_0 - \sum_{m=1}^M {\bs q}_n^{(m)}\right) \right]  \, ,
\end{array}\label{A}
\end{equation}
where $V$ is the volume of the system in the coordinate space.
Making the Laplace transformation, we determine the unnormalized
distribution function
\begin{equation}
\widetilde {\mathbb{F}} _{2N} (\beta ; \bs k_1, \ldots, \bs
k_{2N})  = \int_{E_{\rm min}}^\infty \! \! dE_{\rm tot} e^{
-\beta E_{\rm tot} } \widetilde{f}_{2N}(E_{\rm tot};{\bs k}_1,
\ldots ,{\bs k}_{2N} ) \, . \label{a-12-prime}
\end{equation}
The distribution function in the canonical ensemble reads
${\mathbb{F}}_{2N}(\beta) \! =\widetilde {\mathbb{F}}_{2N}(\beta)/
{Z_{2N}(\beta)}$, where $ {Z_{2N}(\beta)}$ is the partition
function. Skipping over details, we write the result
\begin{equation}
{\mathbb{F}}_{2N}(\beta,{\bs k}_1,\ldots ,{\bs k}_{2N} )
=
\prod_{n=1}^N  f_a(\bs k_n) \, \prod_{n=N+1}^{2N}  f_b(\bs k_n)
\, , \label{Ffact}
\end{equation}
where
\begin{equation}
f_{a(b)}(\bs k)
=
\frac 1{z_{a(b)}(\beta)} \, e^{ -\beta \omega(\bs k) } \, I_M(\bs
k \mp {\bs k}_0) \,, \quad z_{a(b)}(\beta)
=
\int \frac{d^3k}{(2\pi)^3} \, e^{ -\beta \omega(\bs k) } \,
I_M(\bs k \mp{\bs k}_0) \,
 \label{DF}
\end{equation}
are the single-particle distribution functions attributed to
nucleus ``A'' for subindex $a$ or to nucleus ``B'' for subindex
$b$, respectively. Note that $z_a(\beta) =
z_b(\beta) = z(\beta)$ for identical nuclei.
We define the ``multi-scattering form-factor''
\begin{equation}
I_M(\bs Q) \equiv \, \frac 1V \int d^3r \, e^{-i \bs Q \cdot \bs r
} \, \big[ J( {\bs r} )\big]^M  \quad {\mbox{with}} \quad J(\bs
r)=\int \frac{d^3q}{(2\pi)^3} \, J(\bs q) \, e^{i{\bs q \cdot \bs
r} } \, . \label{defi}
\end{equation}
For a large enough number $M$ of the effective collisions and for
the form-factor which possesses the spherical symmetry, $J(\bs q)=
J(|\bs q|)$, we can calculate the first integral in (\ref{defi})
within the saddle-point method
\begin{equation}
I_M(\bs Q) \approx \left( \frac{6\pi} { M \langle q^2 \rangle }
\right)^{3/2} e^{ - \frac{3Q^2} {2M \langle q^2 \rangle}} \, ,
\quad {\rm where}  \quad \langle q^2 \rangle = \int
\frac{d^3q}{(2\pi)^3} \, {\bs q}^2 \, J({\bf q}) \,.
\label{i-largeM}
\end{equation}
In the limit case $M \to \infty$, the dependence on the initial momentum
$\bs k_0$ is washed out, and both single-particle distributions
$f_{a(b)}(\bs k)$ take the same ``thermal'' limit:
$f_{a(b)}(\bs k)  \rightarrow f_{\rm therm}(\bs k) =
 e^{ -\beta \omega(\bs k)}/z_{\rm therm}(\beta)$, where
 $z_{\rm therm}(\beta) = \int d^3k\, e^{ -\beta \omega(\bs k)}$.

Using a ``two-source'' single-particle
distribution function $ f(\bs k_a,\bs k_b)= f_a(\bs k_a)\,
f_b(\bs k_b)$ to average the quantity
$ W(\bs k_a,\bs k_b) = \frac 12 \left[ \delta^3({\bs p}-{\bs k}_a)+
\delta^3({\bs p}-{\bs k}_b) \right]$,
we obtain
\begin{equation}
D_M(\bs p) = \left(\frac 1{2N} \frac{d N}{d^3p} \right)_M =
\frac 1{2z(\beta)} e^{-\beta \omega(\bs p)} \left[ I_M(\bs p -
{\bs k}_0) + I_M(\bs p + {\bs k}_0) \right]. \label{sps}
\end{equation}
It is evident that the spectrum has two items which can be attributed
to the first and second colliding nuclei, respectively, and hence it
can be named a ``two-source single-particle spectrum''.

\vspace{1mm}

\noindent {\bf{Nucleon Rapidity Distribution and Transverse
Spectrum.}}\hspace{2mm}
To obtain the transverse mass and rapidity
distributions, we pass to new variables: $m_\bot=(m^2+{\bs
p}_\bot^2)^{1/2}$, \quad ${\bs p}_{\bot }^{2} =
p_{x}^{2}+p_{y}^{2}$, \, $\dis \tanh{y}= p_z/\omega_p$,
then $d^3p=d\phi \, \omega_p \, m_\bot \, dm_\bot \, dy$.
In accordance with (\ref{model2}),
\begin{equation}
\frac{d^{\,2} N}{ m_\bot dm_\bot dy}
= 2\pi\, m_\bot \cosh y \left[\sum_{M=1}^{M_{\rm max}} C(M) D_M(m_\bot,y)
+ \, C_{\rm therm} \, D_{\rm therm}(m_\bot,y) \right] \, .
\label{tr-total}
\end{equation}
We can define the distribution functions in the rapidity space as
\begin{equation}
\Phi_M(y) \! \equiv \pi \cosh{y} \! \int_{m}^\infty \!
dm_{\perp} \, m_{\perp}^2
\big[ \, f_a(m_\bot, y) + f_b(m_\bot, y)  \, \big] \quad
{\rm with} \quad \int\, dy \, \Phi_M(y) = 1 \, .
\label{phiM}
\end{equation}
Then, the rapidity distribution looks like
\begin{equation}
\frac{dN}{dy}= \sum_{M=1}^{M_{\rm max}} \,C(M) \, \Phi_M(y)
+C_{\rm therm} \,\Phi_{\rm therm}(y) \,.\label{y-fit}
\end{equation}
%

\noindent {\bf{Toy Model.}}\hspace{2mm}
To present the explicit results of our approach, we consider a toy
model: the form-factor $J(\bs q)$ is chosen as a homogeneous
distribution in the sphere of finite radius $q_{\rm max}$,
\begin{equation}
 J({\bs q})
=
\frac{(2\pi)^3}{V_q} \,\, \theta(q_{\rm max}-|\bs q \,|) \, , \ \
\ V_q= \frac 43 \, \pi \, q_{\rm max}^3  \quad {\rm with} \quad
\int \frac{d^3q}{(2\pi)^3} \,J({\bs q})=1 \, . \label{toym}
\end{equation}

\begin{figure}
\includegraphics[width=0.48\textwidth]{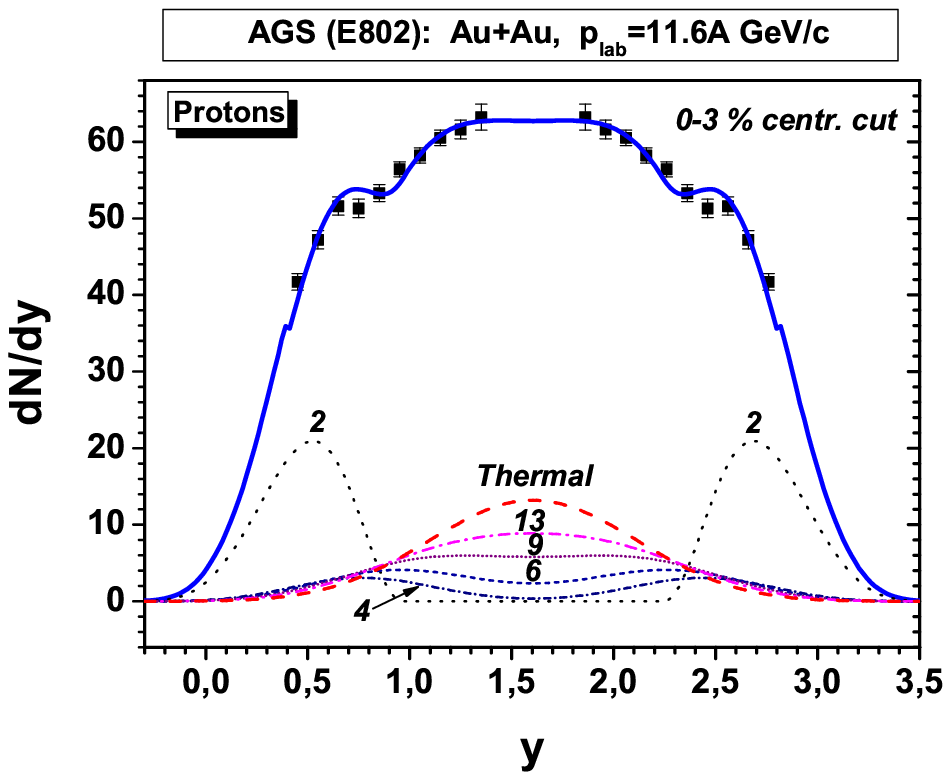}
\hfill
\includegraphics[width=0.48\textwidth]{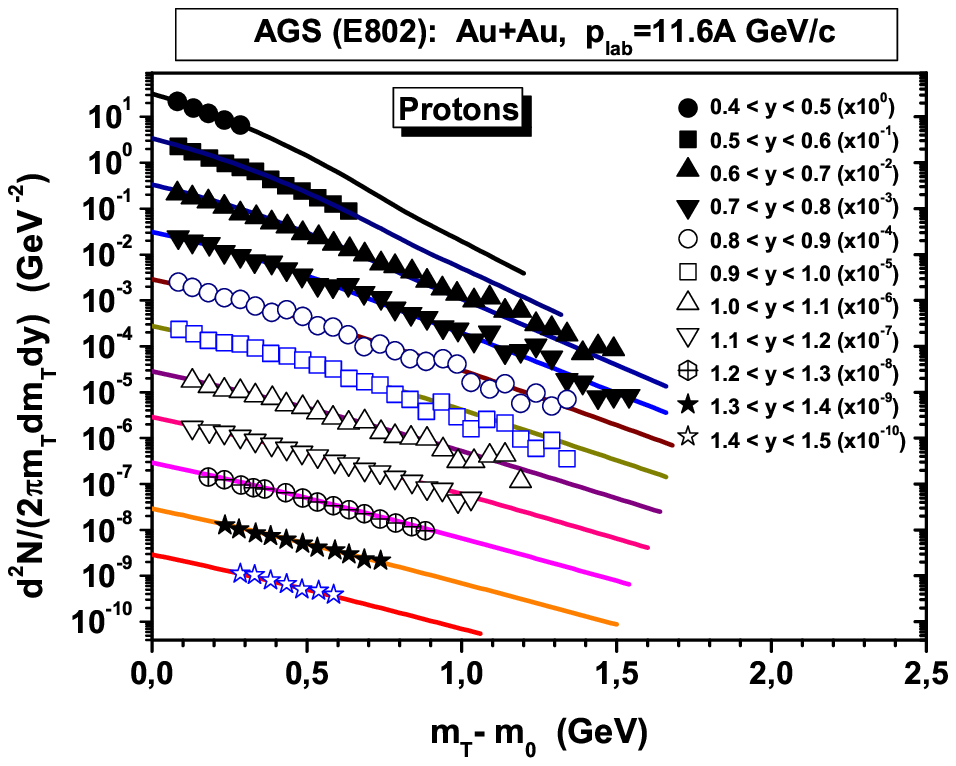}
\\
\parbox[t]{0.48\textwidth}
{\vspace{-7mm} \caption {The result of a fit (thick solid curve)
to experimental data
  \cite{E802-PRC-v60-064901-1999}
  on the rapidity distribution.
  Broken curves marked by numbers $M$ represent the partial
  contributions from every collision ensemble,
  $C(M)\, \Phi_M(y)$
  .}
\label{fig:2a}}
\hfill
\parbox[t]{0.48\textwidth}{\vspace{-7mm} \caption {
Solid curves  represent $m_{\perp}$-spectra obtained in accordance
  with (\ref{tr-total}). We use values of the
  coefficients $C(M)$ which were obtained as a result of the fit to
  the $dN/dy$ data.
  Experimental points are from \cite{E802-PRC-v60-064901-1999}.}
\label{fig:2b}}
\end{figure}

In the proposed  model, the maximum number of collisions, $M_{\rm
max}$, is assumed to be finite and determined by the nucleus number
$A$, initial energy, and centrality.
With the help of the UrQMD transport model \cite{urqmd1,urqmd2}, it was
found that, under the AGS conditions \cite{E802-PRC-v60-064901-1999} with a
centrality of (0-3)\%,
$M_{\rm max}=13$ and $q_{\rm max} = 0.8$~GeV/c (for the toy model
$\frac 35 \, q_{\rm max}^2 = \langle \bs q^2 \rangle =
\langle \bs q^2 \rangle_{_{\rm UrQMD}}$).
Utilizing the thermal distribution, we extract a slope parameter from
experimental data on the proton $m_{\perp}$-spectra
\cite{E802-PRC-v60-064901-1999}, $T=280$~MeV.
Note, the proton data are of interest, first of all, because we know an
exact value of the initial nucleon momentum.
\vspace{-3mm}
\begin{center}
\noindent\begin{tabular}{c c c c c c c c c c c c c c c}
&&&&&&&&&&&&& Table 1 \\
\hline%
   \multicolumn{1}{|c}{$C(1)$} &
   \multicolumn{1}{|c}{$C(2)$} & \multicolumn{1}{|c}{$C(3)$}
 & \multicolumn{1}{|c}{$C(4)$} & \multicolumn{1}{|c}{$C(5)$}
 & \multicolumn{1}{|c}{$C(6)$} & \multicolumn{1}{|c}{$C(7)$}
 & \multicolumn{1}{|c}{$C(8)$} & \multicolumn{1}{|c}{$C(9)$}
 &   \multicolumn{1}{|c}{$C(10)$}
 & \multicolumn{1}{|c}{$C(11)$} & \multicolumn{1}{|c}{$C(12)$}
 & \multicolumn{1}{|c}{$C(13)$} & \multicolumn{1}{|c|}{ $C_{\rm therm}$ }
 \\%
\hline%
   \multicolumn{1}{|c}{1.3} &
   \multicolumn{1}{|c}{23.2} & \multicolumn{1}{|c}{4.7}
 & \multicolumn{1}{|c} {5.3} & \multicolumn{1}{|c} {6.6}
 & \multicolumn{1}{|c} {8.2} & \multicolumn{1}{|c} {9.5}
 & \multicolumn{1}{|c}{10.6}  & \multicolumn{1}{|c}{11.6}
 & \multicolumn{1}{|c}{12.3}
 & \multicolumn{1}{|c}{12.9} & \multicolumn{1}{|c}{13.5}
 & \multicolumn{1}{|c}{13.9} & \multicolumn{1}{|c|}{ 16.6  }
 \\%
\hline
\end{tabular}
\end{center}
%
\vspace{-2mm}
The results of the fit to experimental data
\cite{E802-PRC-v60-064901-1999} on the rapidity distribution and
$m_{\perp}$-spectra of
the net protons are depicted in Figs.~\ref{fig:2a}, \ref{fig:2b}.
The fit was done with making use of expansion (\ref{y-fit})
and resulted in a good description of the experimental data.
It turns out that, in the case of a small number of experimental
points, the set of functions $\Phi_M(y)$ is overcomplete.
To choose a unique configuration of the variable parameters $C(M)$, we
use the maximum entropy method \cite{papoulis}.
%
%
The set of coefficients $C(M)$ (see Table~1) are nothing more as the
absolute number of protons in every collision ensemble.
This description was carried out with account for the contribution of
the thermal source, which cannot appear due to the nucleon-nucleon
(-hadron) rescattering (see the partial contributions in Fig.~1).
We assume that this source is a thermalized multiparton system (QGP)
which emits totally thermalized nucleons through the hadronization
process.
The knowledge of the number of protons, $C_{\rm therm}$, which come
from the QGP, gives us a possibility to evaluate the ``nucleon power''
of the QGP, $P_{\rm QGP}^{(N)}$, created in a particular experiment
on the nucleus-nucleus collision.
We find that, under the AGS conditions \cite{E802-PRC-v60-064901-1999}
(a centrality of 0-3\%), $P_{\rm QGP}^{(N)}\approx 11\%$.
So, in the framework of the proposed criterion, it could be claimed
that the QGP (as a nucleon source) was created not only at
SPS energies \cite{heinz2000}, but it was also created in the
central collisions at AGS energies.

Meanwhile, the partial expansion,
$dN/dy=\sum_{M=1}^{M_{\rm max}}C(M) \, \varphi_M(y)$
(see (\ref{y-fit})), where we do not use the thermal contribution,
makes a good description (the same $\chi^2$) of the experimental data
on both the rapidity distribution and $m_{\perp}$-spectra.
So, we cannot resolve unambiguously the presence of the thermal
source.
In fact, to overcome the problem, we need a more detailed experimental
information for the central rapidity region.

All this encourages us to apply the model to other experiments and
problems.


\vspace{-4mm}

\begin{thebibliography}{50}
\vspace{-1mm}
%
\bibitem{urqmd1}
S. A. Bass, M. Belkacem, M. Bleicher et al.,
Prog. Part. Nucl. Phys. {\bf 41},  225 (1998).

\bibitem{urqmd2}
M. Bleicher, E. Zabrodin, C. Spieles et al.,
J. Phys. G: Nucl. Part. Phys. {\bf 25}, 1859 (1999).

\bibitem{anch-2008-1}
D.~Anchishkin, S.~Yezhov, ArXiv:\,0802.0259 [nucl-th].

\bibitem{E802-PRC-v60-064901-1999}
L. Ahle et al. (E802 Collaboration), Phys. Rev. C {\bf 60}, 064901
(1999).

\bibitem{papoulis} A. Papoulis, Probability, Random Variables and Stochastic
Processes, McGraw-Hill, NY 2002.

\bibitem{heinz2000}
Ulrich W. Heinz, Maurice Jacob, ArXiv:nucl-th/0002042.


\end{thebibliography}
\end{document}